\documentclass[sn-mathphys]{sn-jnl}

\jyear{2024}%

\theoremstyle{thmstyleone}%
\theoremstyle{thmstyletwo}%

\theoremstyle{thmstylethree}%

\begin{document}

\title[Three-dimensional integral Faddeev equations without a certain symmetry ]{Three-dimensional integral Faddeev equations without a certain symmetry}


\author*[1]{\fnm{Egorov} \sur{Mikhail}}\email{egorovphys@mail.com, https://orcid.org/0000-0001-8231-3884}



\affil[1]{\orgdiv{Physics Department}, \orgname{Tomsk State University}, \orgaddress{ \city{Tomsk},  \country{Russia}}}
\affil*{\orgdiv{Bogolyubov Laboratory of Theoretical Physics}, \orgname{Joint Institute for Nuclear Research}, \orgaddress{ \city{Dubna},  \country{Russia}}}



\abstract{
	
 A method for the direct integration of the three-dimensional Faddeev equations with respect to the breakup T-matrix in momentum space for three-body systems with differing masses is presented. The Faddeev equations are explicitly formulated without imposing symmetry or antisymmetry requirements on the two-body t-matrices, thus accounting for mass differences between the three interacting particles. An algorithm for the algebraic determination of non-relativistic wave functions for three-body systems with arbitrary masses is given. Furthermore, it is directly demonstrated how the domain of logarithmic singularities in the integral kernels of the Faddeev equations is significantly altered by varying the masses of the interacting particles.
	
}




\maketitle

\section{Introduction}\label{sec1}
 
The solutions to the dynamic equations of Faddeev \cite{Fadd} and Faddeev-Yakubovsky \cite{FaddYakub}, along with numerous generalizations and simplifications (see, e.g., \cite{AGS67,AGS69,Mukh12,Mukh18} and also review of four-nucleon calculations \cite{Fon17}), provided a strong foundation for precise methods in the nonrelativistic theory of interacting three- and few-body systems, respectively. The Faddeev and Faddeev-Yakubovsky equations are best known in their separable form, also known as the AGS form \cite{AGS70}, the applicability of which is ensured by the rapid convergence of the Hilbert-Schmidt norm with increasing scattering energy. 
Few-body dynamics in the AGS form have proven themselves well not only in the description of elastic scattering and nuclear reactions at low energies, but also in various types of eigenvalue problems with the search for binding energies in exotic meson-nuclear \cite{Belyaev06,Shevchenko07,Shevchenko21} and hyperon-nuclear \cite{Nogga02,Nogga13,Egorov23} systems.  

Methods for accurately accounting for the Coulomb interaction in the few-body dynamics of arbitrary charged, strongly interacting particles within the Faddeev approach in momentum space are also well-known \cite{Mukh00,Mukh01,Deltuva05a,Deltuva05b,Deltuva08,Deltuva19}. Two-potential methods for treating the Coulomb interaction in three-body dynamics with Coulomb  off shell effects have also been developed \cite{Oryu06,Oryu06E}. Calculations of Coulomb effects using the two-potential method are known to be independent on the choice of the screening region for the Coulomb potential and do not also require the use of regularization techniques.

One area of few-body physics that remains relatively undeveloped is the solution of the Faddeev equations in their integral form, without resorting to partial wave decomposition \cite{Gloeckle96,Elster99,Liu05}. Direct numerical integration of the three-dimensional equations for the breakup T-matrix has remained underutilized, largely due to the computational demands on early computers and the traditional reliance on low-order partial wave expansions for the short-range potentials commonly used in few-body calculations (see, e.g., \cite{Stingl,Kharchenko80}). However, advances in computer technology and the growing application of precise few-body methods in atomic-molecular systems \cite{Kilic04,Kolganova11,Frolov19,Kolganova19}  are creating new demands. This includes both expanding the types of potentials used and requiring improvements in computational methods and the approximations employed, which were often specific to short-range interactions. Direct numerical integration of the Faddeev equations, without partial wave decomposition, has only been applied to scattering problems in the simplest, symmetrized case of nucleon scattering on a coupled system—the deuteron \cite{Liu05}. This work demonstrated the viability of the numerical methods and achieved good agreement between the results and experimental data.

When considering the complex energy plane to explore virtual and resonant three-body states, one encounters significant challenges in multichannel dynamics where the number of coupled dynamic equations increases with changes in the particle types. Given that two-body interactions must be analytically continued into the unphysical energy sheets for each partial wave, as done in Ref. \cite{Orlov81}, further increasing the number of coupled equations via partial wave expansions ceases to be a simplifying approach. Even without changing the topology of the energy surface, the inclusion of Coulomb interactions requires justification through a truncation of the lower partial waves. Consequently, direct numerical integration for the scattering matrix becomes an increasingly attractive method, both for solving scattering problems and for addressing spectral problems that involve analytical continuation of two-body potentials, as well as two- and three-body t-matrices, onto unphysical energy sheets \cite{Orlov81,Orlov84}.

This work focuses specifically on the influence of mass differences between interacting particles as they appear in the three-dimensional integral Faddeev equations, which results in eigenfunctions lacking specific symmetry properties. It is explicitly derived the matrix elements of all particle permutation operators in the vector basis, and   their action on the momentum dependence of the two-body t-matrices and the three-body system eigenfunctions is also demonstrated. Once formulation of the integral Faddeev equations in vector form is used to provide expressions for the eigenvalue wave functions of three-body systems,  the calculations of these functions for the $^3$He system is also presented. One analyzes in detail how the domain of logarithmic singularities arising in the integral kernels of the inhomogeneous Faddeev equations changes as the masses of the interacting particles are varied over a wide range.

	\section{Equations for the three-body breakup T-matrix without certain symmetry} \vspace{5mm}

For three bodies of different masses, the Jacobi variables $\vec p_i$, $\vec q_i$, $i\in[1,2,3]$ where the index $i$-enumerates the particles for which $\vec p$ is the relative momentum in the interacting pair $i$, and $\vec q$-the spectator momentum of the particle $i$ relative to this pair will have two equivalent representations expressing these momenta through each other
\begin{align}\label{eq1}
	\begin{gathered}
		\vec p_1=-\frac{m_2}{m_2+m_3}\vec p_2+\frac{\vec q_2}{m_2+m_3}\Big(m_3+\frac{m_2m_3}{m_3+m_1}\Big);
		\vec q_1=-\vec p_2-\vec q_2\frac{m_1}{m_1+m_3};\\
		\vec p_1=-\frac{m_3}{m_2+m_3}\vec p_3-\vec q_3\frac{m_2M}{(m_2+m_3)(m_1+m_2)}; \vec q_1=\vec p_3-\vec q_3 \frac{m_1}{m_1+m_2}; \\
		\vec p_2=-\frac{m_1}{m_3+m_1}\vec p_1-\frac{q_1}{m_3+m_1}\Big(\frac{m_1m_3}{m_2+m_3}+m_3\Big); 
		\vec q_2=\vec p_1-\vec q_1\frac{m_2}{m_2+m_3}; \\
		\vec p_2=-\frac{m_3}{m_3+m_1}\vec p_3+\frac{\vec q_3}{m_3+m_1}\Big(m_1+\frac{m_1m_3}{m_1+m_2}\Big); 
		\vec q_2=-\vec p_2-\vec q_3\frac{m_2}{m_1+m_2}; \\
		\vec p_3=-\frac{m_1}{m_1+m_2}\vec p_1+\frac{\vec q_1}{m_1+m_2}\Big(m_2+\frac{m_1m_2}{m_2+m_3}\Big); 
		\vec q_3=-\vec p_1-\vec q_1\frac{m_3}{m_2+m_3}; \\
		\vec p_3=-\frac{m_2}{m_1+m_2}\vec p_2-\frac{\vec q_2}{m_1+m_2}\Big(\frac{m_2m_1}{m_1+m_3}+m_1\Big);
		\vec q_3=\vec p_2-\vec q_2\frac{m_3}{m_3+m_1}. 
	\end{gathered}
\end{align}
Here $M=m_1+m_2+m_3$ is the sum of the masses of the three particles. 
Two-body $t$ matrices are known enter the three-body phase space together with delta functions
that exclude the dynamic presence of a third particle in two-body domain. In the $\vec p$, $\vec q$ representation for some vector $\mid\Psi'\rangle\equiv\mid tP\Psi\rangle$, where $\mid\Psi\rangle$ is a stationary state of a three-body system, one should has
\begin{align}\label{eq2}
	\begin{gathered}
		\langle\vec q,\vec p\mid tP\Psi\rangle=\int d^3q'd^3p'd^3q^{''}d^3p^{''}\langle\vec p,\vec q\mid t\mid \vec p',\vec q'\rangle\langle\vec p',\vec q'\mid P\mid\vec p^{''},\vec q^{''}\rangle\langle\vec p^{''},\vec q^{''}\mid \Psi\rangle=\\
		=\int d^3q'd^3p'd^3q^{''}d^3p^{''}\langle\vec p\mid t\mid \vec p'\rangle\delta^{(3)}(\vec q-\vec q')\Big(\delta^{(3)}(\vec p'-\vec p^{''})\delta^{(3)}(\vec q'-\vec q^{''})\Big)\langle\vec p^{''},\vec q^{''}\mid\Psi\rangle.	
	\end{gathered}	
\end{align}	
However, the dependence on relative  momenta in interacting pairs of particles is significantly distorted due to the presence of the integral kernel of the permutation operator $P$-two three-dimensional delta functions. 
For simplicity, the partition indices in (\ref{eq2}) are suppressed. 
The introduction of a permutation operator, $P$, within an approach lacking a definite symmetry, i.e., for three distinct particles, might seem peculiar at first glance. While  there are three equivalent sets of Jacobi coordinates (\ref{eq1}) for a given partition,  one must also account for the two ways to express the pair of momenta within that fixed partition which are needed for the $t$--matrix calculations in terms of the other two pairs of momenta that define the initial state. 
 This crucial point implies that the right-hand side of equation (\ref{eq2}) actually involves two terms, each corresponding to one of the two ways to express the momenta $\vec p'_i, \vec q'_i$ in terms of the momenta $\vec p^{''}_j,\vec q^{''}_j$ for $j\not=i$.  Consequently, the delta-functions remove the integrations in a non-trivial manner, leading to a modification of the momentum dependence of the $t$-matrix via its momentum $\vec p'$ and the wave function $\langle\vec p^{''},\vec q^{''}\mid\Psi\rangle $ via its momentum  $\vec p^{''}$.  

Removing six integrals in expressions (\ref{eq2}) using kinematic relations (\ref{eq1}) one leads to the following explicit representations for two-body $t$ matrices
 \begin{align}\label{eq3}
	\begin{gathered}
		\langle\vec q,\vec p\mid t_1P\Psi\rangle=
		\int d^3q^{''}\langle \vec p\mid t_1\mid -\vec q \tfrac{m_1}{m_1+m_3}-\vec q^{''}\rangle\langle\vec q+\vec q^{''}\tfrac{m_2}{m_2+m_3},\vec q^{''}\mid\Psi\rangle+ \\
		+\langle\vec p\mid t_1\mid \vec q\tfrac{m_1}{m_1+m_2}+\vec q^{''}\rangle\langle-\vec q-\vec q^{''}\tfrac{m_3}{m_2+m_3},\vec q^{''}\mid\Psi\rangle; \\
		\langle\vec q,\vec p\mid t_2P\Psi\rangle=
		\int d^3q^{''}\langle \vec p\mid t_2\mid \vec q \tfrac{m_2}{m_2+m_3}+\vec q^{''}\rangle\langle-\vec q-\vec q^{''}\tfrac{m_1}{m_3+m_1},\vec q^{''}\mid\Psi\rangle+ \\
		+\langle\vec p\mid t_2\mid-\vec q\tfrac{m_2}{m_1+m_2}-\vec q^{''}\rangle\langle\vec q+\vec q^{''}\tfrac{m_3}{m_3+m_1},\vec q^{''}\mid\Psi\rangle;  \\
		\langle\vec q,\vec p\mid t_3P\Psi\rangle=
		\int d^3q^{''}\langle \vec p\mid t_3\mid-\vec q \tfrac{m_3}{m_2+m_3}-\vec q^{''}\rangle\langle\vec q+\vec q^{''}\tfrac{m_1}{m_1+m_2},\vec q^{''}\mid\Psi\rangle+ \\
		+\langle\vec p\mid t_3\mid\vec q\tfrac{m_3}{m_3+m_1}+\vec q^{''}\rangle\langle-\vec q-\vec q^{''}\tfrac{m_2}{m_1+m_2},\vec q^{''}\mid\Psi\rangle. 
	\end{gathered}
\end{align}

The system of  coupled Faddeev equations for a three-body $T_i$ matrix, where the index i-characterizes a spectator particle marked relative to an interacting pair, without introducing a certain symmetry with respect to the permutation of particles in places, is represented as
\begin{align}\label{eq4}
	\begin{gathered}	
		T_1=t_1\Psi_2+t_1\Psi_3+t_1R_0T_2+t_1R_0T_3; \\	
		T_2=t_2\Psi_1+t_2\Psi_3+t_2R_0T_1+t_2R_0T_3; \\
		T_3=t_3\Psi_1+t_3\Psi_2+t_3R_0T_1+t_3R_0T_2, 		
	\end{gathered}
\end{align}
where $R_0$ is an interaction-free three-body Green function.

 Choosing a coordinate system according to work \cite{Liu05}, in which the momentum $\vec q_0$ characterizes the relative momentum of a particle impinging on a bound system. Then in the selected $\vec p$, $\vec q$ representation, in which
$\langle\vec p,\vec q\mid T\Psi\rangle\equiv 
T(p,x_p,x_{pq}^{q_0},x_q,q;q_0)\langle\vec p,\vec q\mid\Psi\rangle$
and $x_p$, $x_q$ are the cosines of the polar angles  of the vectors $\vec p$ and $\vec q$, respectively, and $x_{pq}^{q_0}$ is the angle between the normals drawn for the planes $\vec p-\vec q_0$ and $\vec q-\vec q_0$, the system (\ref{eq4}) takes the following final form
\begin{align}\label{eq5}
	\begin{gathered}
		T_1(p,x_p,x_{pq}^{q_0},x_q,q)=t_1\Big(p,p_{11}',\theta_{pp_{11}'}\Big)
		\langle\vec q+\vec q_0\tfrac{m_2}{m_2+m_3},\vec q_0\mid\Psi\rangle \\
		+t_1\Big(p,p_{12}',\theta_{pp_{12}'}\Big)\langle-\vec q-\vec q_0\tfrac{m_3}{m_2+m_3},\vec q_0\mid\Psi\rangle 
		+\int d^3q^{''}\Big[ t_1\Big(p,f_{11}(q^{''}),\theta_{f_{11}q''}\Big)\\
		\Big(E-\tfrac{q^2}{2\mu_{23}}-\tfrac{q^{''2}}{2\mu_{13}}-\tfrac{qq^{''}y_{qq^{''}}}{m_3}\Big)^{-1}
		T_2\Big(g_{11}(q^{''}),X_{g_{11}},x_{g_{11} q^{''}}^{q_0},x_{q^{''}},q^{''}\Big) \\
		+t_1\Big(p,f_{12}(q^{''}),\theta_{f_{12}q^{''}}\Big) 
		\Big(E-\tfrac{q^2}{2\mu_{23}}-\tfrac{q^{''2}}{2\mu_{12}}-\tfrac{qq^{''}y_{qq^{''}}}{m_2}\Big)^{-1}\\
		T_3\Big(g_{12}(q^{''}),X_{g_{12}},x_{g_{12}q^{''}}^{q_0},x_{q^{''}},q^{''}\Big)\Big]; \\
		T_2(p,x_p,x_{pq}^{q_0},x_q,q)=t_2\Big(p,p_{21}',\theta_{pp_{21}'}\Big)
		\langle-\vec q-\vec q_0\tfrac{m_1}{m_3+m_1},\vec q_0\mid\Psi\rangle \\
		+t_2\Big(p,p_{22}',\theta_{pp_{22}'}\Big)\langle\vec q+\vec q_0\tfrac{m_3}{m_3+m_1},\vec q_0\mid\Psi\rangle 
		+\int d^3q^{''}\Big[ t_2\Big(p,f_{21}(q^{''}),\theta_{f_{21}q''}\Big) \\
		\Big(E-\tfrac{q^2}{2\mu_{31}}-\tfrac{q^{''2}}{2\mu_{23}}-\tfrac{qq^{''}y_{qq^{''}}}{m_3}\Big)^{-1}
		T_1\Big(g_{21}(q^{''}),X_{g_{21}},x_{g_{21} q^{''}}^{q_0},x_{q^{''}},q^{''}\Big) \\
		+t_2\Big(p,f_{22}(q^{''}),\theta_{f_{22}q^{''}}\Big)\Big(E-\tfrac{q^2}{2\mu_{31}}-\tfrac{q^{''2}}{2\mu_{12}}-\tfrac{qq^{''}y_{qq^{''}}}{m_1}\Big)^{-1}\\
		T_3\Big(g_{22}(q^{''}),X_{g_{22}},x_{g_{22}q^{''}}^{q_0},x_{q^{''}},q^{''}\Big)\Big]; \\
		T_3(p,x_p,x_{pq}^{q_0},x_q,q)=t_3\Big(p,p_{31}',\theta_{pp_{31}'}\Big)
		\langle\vec q+\vec q_0\tfrac{m_1}{m_1+m_2},\vec q_0\mid\Psi\rangle \\
		+t_3\Big(p,p_{32}',\theta_{pp_{32}'}\Big)\langle-\vec q-\vec q_0\tfrac{m_2}{m_1+m_2},\vec q_0\mid\Psi\rangle 
		+\int d^3q^{''}\Big[ t_3\Big(p,f_{31}(q^{''}),\theta_{f_{31}q''}\Big) \\
		\Big(E-\tfrac{q^2}{2\mu_{12}}-\tfrac{q^{''2}}{2\mu_{23}}-\tfrac{qq^{''}y_{qq^{''}}}{m_2}\Big)^{-1}
		T_1\Big(g_{31}(q^{''}),X_{g_{31}},x_{g_{31} q^{''}}^{q_0},x_{q^{''}},q^{''}\Big) \\
		+t_3\Big(p,f_{32}(q^{''}),\theta_{f_{32}q^{''}}\Big)\Big(E-\tfrac{q^2}{2\mu_{12}}-\tfrac{q^{''2}}{2\mu_{31}}-\tfrac{qq^{''}y_{qq^{''}}}{m_1}\Big)^{-1}\\
		T_2\Big(g_{32}(q^{''}),X_{g_{32}},x_{g_{32}q^{''}}^{q_0},x_{q^{''}},q^{''}\Big)\Big]; 
	\end{gathered}
\end{align}   
where, for simplicity, the parametric dependence of all $T$ matrices on the momentum $q_0$ was omitted, and the values $\mu_{i,j\not=i}$ are the usual reduced particle masses. The notation is introduced in (\ref{eq5}) for convenience
\begin{align}\label{eq6}
	\begin{gathered}
		f_{11}(q^{''})=\sqrt{\Big(q\frac{m_1}{m_1+m_3}\Big)^2+q^{''2}+2qq^{''}\frac{m_1}{m_1+m_3}y_{qq^{''}}}; \\
		\theta_{f_{11}q^{''}}=-y_{pq}\frac{q}{f_{11}(q^{''})}\frac{m_1}{m_1+m_3}-y_{pq^{''}}\frac{q^{''}}{f_{11}(q^{''})}; \\
		g_{11}(q'')=\sqrt{q^2+\Big(q^{''}\frac{m_2}{m_2+m_3}\Big)^2+2qq^{''}\frac{m_2}{m_2+m_3}y_{qq^{''}}}; \\
		X_{g_{11}}=y_{qq_0}\frac{q}{g_{11}(q^{''})}+y_{q^{''}q_0}\frac{q^{''}}{g_{11}(q^{''})}\frac{m_2}{m_2+m_3};  \\
		x_{g_{11}q^{''}}^{q_0}=
		\frac{y_{qq^{''}}\tfrac{q}{g_1(q^{''})}+\tfrac{q^{''}}{g_1(q^{''})}\tfrac{m_2}{m_2+m_3}-a_{qq^{''}}x_{q^{''}}}
		{\sqrt{1-a_{qq^{''}}^2}\sqrt{1-x_{q^{''}}^2}}; \\
		a_{qq^{''}}=y_{qq_0}\frac{q}{g_{11}(q^{''})}+y_{q^{''}q_0}\frac{q^{''}}{g_{11}(q^{''})}\frac{m_2}{m_2+m_3}; 	 \\
		f_{12}(q^{''})=\sqrt{\Big(q\frac{m_1}{m_1+m_2}\Big)^2+q^{''2}+2qq^{''}\frac{m_1}{m_1+m_2}y_{qq^{''}}}; \\
		\theta_{f_{12}q^{''}}=y_{pq}\frac{q}{f_{12}(q^{''})}\frac{m_1}{m_1+m_2}+y_{pq^{''}}\frac{q^{''}}{f_{12}(q^{''})}; \\
		g_{12}(q'')=\sqrt{q^2+\Big(q^{''}\frac{m_3}{m_2+m_3}\Big)^2+2qq^{''}\frac{m_3}{m_2+m_3}y_{qq^{''}}}; \\	
		X_{g_{12}}=-y_{qq_0}\frac{q}{g_{12}(q^{''})}-y_{q^{''}q_0}\frac{q^{''}}{g_{12}(q^{''})}\frac{m_3}{m_2+m_3};  \\
		x_{g_{12}q^{''}}^{q_0}=
		\frac{-y_{qq^{''}}\tfrac{q}{g_2(q^{''})}-\tfrac{q^{''}}{g_2(q^{''})}\tfrac{m_3}{m_2+m_3}-b_{qq^{''}}x_{q^{''}}}
		{\sqrt{1-b_{qq^{''}}^2}\sqrt{1-x_{q^{''}}^2}}; \\
		b_{qq^{''}}=-y_{qq_0}\frac{q}{g_{12}(q^{''})}-y_{q^{''}q_0}\frac{q^{''}}{g_{12}(q^{''})}\frac{m_3}{m_2+m_3}.	 
	\end{gathered}	
\end{align}	
and also
\begin{align}\label{eq7}
	\begin{gathered}
		p_{11}'=\sqrt{\Big(q\frac{m_1}{m_1+m_3}\Big)^2+q_0^2+2qq_0\frac{m_1}{m_1+m_3}y_{qq_0}};  \\
		p_{12}'=\sqrt{\Big(q\frac{m_1}{m_1+m_2}\Big)^2+q_0^2+2qq_0\frac{m_1}{m_1+m_2}y_{qq_0}};   \\
		\theta_{pp_{11}'}=-y_{pq}\frac{q}{p_1'}\frac{m_1}{m_1+m_3}
		-y_{pq_0}\frac{q_0}{p_1'},\,  \theta_{pp_{12}'}=y_{pq}\frac{q}{p_2'}\frac{m_1}{m_1+m_2}+y_{pq_0}\frac{q_0}{p_2'};  \\
		y_{q^{''}q_0}=x^{''}x_{q_0}+\sqrt{1-x^{''2}}\sqrt{1-x_{q_0}^2}\cos{(\phi^{''}-\phi_{q_0})},\,y_{qq_0}=x_q;   \\
		y_{pq^{''}}=y_{pq}x^{''}+\sqrt{1-x^{''2}}\sqrt{1-y_{pq}^2}\cos{(\phi_p-\phi^{''})},\,y_{pq_0}=x_p;  \\
		y_{pq}=x_qx_p+\sqrt{1-x_q^2}\sqrt{1-x_p^2}\cos{(\phi_q-\phi_p)},\,y_{qq^{''}}=x^{''}; \\
		x_{q^{''}}=x^{''}x_{q_0}+\sqrt{1-x^{''2}}\sqrt{1-x_{q_0}^2}\cos{(\phi^{''}-\phi_{q_0})}. 
	\end{gathered}	
\end{align}		
In addition to obtain the expressions $p_{21}'$, $p_{22}'$, $\theta_{pp_{21}'}$, $\theta_{pp_{22}'}$, $f_{21}(q^{''})$, $\theta_{f_{21}q^{''}}$, $f_{22}(q^{''})$, $\theta_{f_{22}q^{''}}$, $g_{21}(q^{''})$, $g_{22}(q^{''})$, $X_{g_{21}}$, $X_{g_{22}}$, $x_{g_{21}q^{''}}^{q_0}$ and $x_{g_{22}q^{''}}^{q_0}$ it is enough in the formulas (\ref{eq6},\ref{eq7}) to replace the masses $m_1\leftrightarrow m_2$ and change the signs at all angles $\theta_{pp_{11}'}$, $\theta_{pp_{12}'}$, $\theta_{f_{11}q^{''}}$, $\theta_{f_{12}q^{''}}$, $X_{g_{11}}$, $X_{g_{12}}$, $x_{g_{11}q^{''}}^{q_0}$ and $x_{g_{12}q^{''}}^{q_0}$ to the opposite.  It can also be shown that the expressions for $p_{31}'$, $p_{32}'$, $\theta_{pp_{31}'}$, $\theta_{pp_{32}'}$, $f_{31}(q^{''})$, $\theta_{f_{31}q^{''}}$, $f_{32}(q^{''})$, $\theta_{f_{32}q^{''}}$, $g_{31}(q^{''})$, $g_{32}(q^{''})$, $X_{g_{31}}$, $X_{g_{32}}$, $x_{g_{31}q^{''}}^{q_0}$ and $x_{g_{32}q^{''}}^{q_0}$ are obtained from the formulas (\ref{eq6},\ref{eq7}) by replacing the masses ($m_1,m_2,m_3$) with ($m_3,m_1,m_2$) with the same sign at all angles.

By setting the three-body scattering energy $E$ and integrating the expression (\ref{eq6}), which includes two-body $t$ matrices defined on half off mass shell, one can obtain a solution of  the  three-body  scattering problem  relative to the breakup  $T$-matrices $T_1$, $T_2$, and $T_3$. Subsequently, these $T$ matrices can be used, as is known \cite{Gloeckle96}, in finding the amplitudes of elastic scattering and reactions.

Having fixed the initial state of the system in the form of $\Psi_{1(23)}$, where the pair (23) forms a coupled state, which is described by its eigenstate function $\phi_{(23)}(\vec q)$, the transition operator will formally act according to the rule
\begin{align}\label{eq8}
\begin{gathered}
	U\mid\Psi_{1(23)}\rangle=(P_{12}P_{23}+P_{13}P_{23})\big(R_0^{-1}+T_1\big)\Psi_{1(23)}=\\
	\big(R_0^{-1}+T_2\big)\Psi_{2(31)}+\big(R_0^{-1}+T_3\big)\Psi_{3(12)}. 	
\end{gathered}
\end{align}	
The expression for the   elastic scattering amplitude $\langle\Psi\mid U\mid\Psi\rangle$ is obtained from the matrix elements of the transition operator and has an explicit form for a system of three different masses
\begin{align}\label{eq9}
	\begin{gathered}
		\langle\Psi_{1(23)}\mid U\mid\Psi_{1(23)}\rangle=\\
		=\phi_{(23)}\big(-\vec q\tfrac{m_1}{m_1+m_3}-\vec q_0\big)\big(E-\tfrac{q^2}{2\mu_{23}}-\tfrac{q_0^2}{2\mu_{13}}-\tfrac{q_0qy_{qq_0}}{m_3}\big)\phi_{(31)}(\vec q+\vec q_0\tfrac{m_2}{m_2+m_3})+ \\
		+\phi_{(23)}\big(\vec q\tfrac{m_1}{m_1+m_2}+\vec q_0\big)\big(E-\tfrac{q^2}{2\mu_{23}}-\tfrac{q_0^2}{2\mu_{12}}-\tfrac{q_0qy_{qq_0}}{m_2}\big)\phi_{(12)}(-\vec q-\vec q_0\tfrac{m_3}{m_2+m_3})+ \\
		+\int d^3q'\Big[\langle\Psi_{1(23)}\mid-\vec q\tfrac{m_1}{m_1+m_3}-\vec q',\vec q'\rangle\langle\vec q+\vec q'\tfrac{m_2}{m_2+m_3},\vec q'\mid T_2\mid\Psi_{2(31)}\rangle+\\
		+\langle\Psi_{1(23)}\mid\vec q\tfrac{m_1}{m_1+m_2}+\vec q',\vec q'\rangle \langle-\vec q-\vec q'\tfrac{m_3}{m_2+m_3},\vec q'\mid T_3\mid\Psi_{3(12)}\rangle\Big].
	\end{gathered}	
\end{align}
It is important to note that in elastic scattering, the free motion of the spectator's particle (not coupled in the pair) is split off and the momentum of the bombarding particle $\vec q_0$ is preserved in magnitude, whereas for scattering with rearrangement, the momentum $\vec q_0$  is equal in modulus to the final particle momentum $\vec q$.  Antisymmetrization of three-body states $\psi_{1(23)}$ for the case of three identical fermions directly leads equation (\ref{eq9}) to a simple case  given in the work \cite{Liu05}.

The breakup amplitude $\langle\vec p,\vec q\mid U_0\mid\Psi_{1(23)}\rangle$ is obtained by the action of the breakup operator $U_0=(1+P)T_1\Psi_{1(23)}$ and  in accepted notation has the form
\begin{align}\label{eq10}
	\begin{gathered}
		\langle\vec p,\vec q\mid U_0\mid\Psi_{1(23)}\rangle=T_1(p,x_p,x_{pq}^{q_0},x_q,q)+
		T_2(p_2,x_{p_2},x_{p_2q_2}^{q_0},x_{q_2},q_2) \\
		+T_3(p_3,x_{p_3},x_{p_3q_3}^{q_0},x_{q_3},q_3).
	\end{gathered}
\end{align}
In the selected frame of reference, when the Oz axis is aligned with the momentum $\vec q_0$, one has for independent variables of $T$ matrices $\langle\vec p',\vec q'\mid T_k\mid\Psi_{k(ij)}\rangle\equiv T_k(p_k',x_{p_k'},x_{p_k'q'}^{q_0},x',q')$ (\ref{eq9}), where $k\in[2,3]$ and $i\not=j\not=k$  the following explicit expressions
\begin{align}\label{eq11}
	\begin{gathered}
		p_k'=\sqrt{q^2+\Big(q'\frac{m_k}{m_2+m_3}\Big)^2+2qq'\frac{m_k}{m_2+m_3}y_{qq'}};\\
		x_{p_k'}=(-1)^{k}x_q\frac{q}{p_k'}+(-1)^kx'\frac{q'}{p_k'}\frac{m_k}{m_2+m_3}x';\\
		x_{p_k'q'}^{q_0}=(-1)^k\frac{y_{qq'}\tfrac{q}{p_k'}+\tfrac{q'}{p_k'}\tfrac{m_k}{m_2+m_3}-c_{qq'}x'}{\sqrt{1-x^{'2}}\sqrt{1-c_{qq'}^2}};\\
		c_{qq'}=(-1)^kx_q\frac{q}{p_k'}+(-1)^kx'\frac{q'}{p_k'}\frac{m_k}{m_2+m_3};\\
		y_{qq'}=x_qx'+\sqrt{1-x_q^2}\sqrt{1-x^{'2}}\cos{(\phi_q-\phi')}.
	\end{gathered}
\end{align} 
Kinematic variables for the breakup $T$-matrices of the  breakup amplitude have a more complex form (\ref{eq10})
 \begin{align}\label{eq12}
	\begin{gathered}
		p_k=\\
		\sqrt{\Big(\frac{pm_k}{m_2+m_3}\Big)^2+\Big(\frac{qm_lM}{(m_2+m_3)(m_1+m_l)}\Big)^2+(-1)^l2pqy_{pq}\frac{m_2m_3M}{(m_2+m_3)^2(m_1+m_l)}};\\
		y_{pq}=x_px_q+\sqrt{1-x_p^2}\sqrt{1-x_q^2}x_{pq}^{q_0};\\
		q_{k}=\sqrt{p^2+\Big(\frac{qm_1}{m_1+m_l}\Big)^2+2pq\frac{m_1}{m_1+m_l}y_{pq}};\\
		x_{p_k}=-\frac{m_k}{m_2+m_3}\frac{p}{p_k}x_p+(-1)^k\frac{q}{p_k}\frac{m_lM}{(m_2+m_3)(m_1+m_l)}x_q;\\
		x_{q_k}=-\frac{p}{q_k}x_p-\frac{q}{q_k}\frac{m_1}{m_1+m_l}x_q; \\
		x_{p_kq_k}^{q_0}=\\
		\frac{\tfrac{(-1)^km_kp^2}{p_kq_q(m_2+m_3)}+\tfrac{m_1m_k-m_lM}{(m_2+m_3)(m_1+m_l)}\frac{pq}{p_kq_k}y_{pq}-\tfrac{(-1)^km_1m_lM}{(m_2+m_3)(m_1+m_l)^2}\tfrac{q^2}{p_kq_k} -x_{p_k}x_{q_k}}{\sqrt{1-x_{p_k}^2}\sqrt{1-x_{q_k}^2}},
	\end{gathered}
\end{align}	
where the indices $k,l\in[2,3]$ run through the same values and $k\not=l$.

\section{Eigenstates of the system}\label{sec2}

It is natural to expect that, as in the case of three identical particles (fermions or bosons), the eigenfunction of the system (\ref{eq5}) is also found as an eigenvector of a homogeneous algebraic system of equations for a given binding energy $E_b$ of three-body system.

Let $a_{ij}$, $i\not=j$, $i,j\in[1,2,3]$ are the integral kernels of a homogeneous system of equations (\ref{eq5}). Then the eigenvector of this system is found by a simple algebraic approximation of the integral equation to a numerical grid of nodes $N\times N$, where $N$ is the number of grid nodes along the momentum $\vec q$ or $\vec q_0$
\begin{align}\label{eq13}
	\begin{gathered}
		\Bigg[\Bigg(
		\begin{matrix}
			1_{N\times N} & 0 & 0\\
			0 & 1_{N \times N} & 0 \\
			0 & 0 & 1_{N\times N} 
		\end{matrix}
		\Bigg)
		-
		\Bigg(
		\begin{matrix}
			0 & a_{12} & a_{13} \\
			a_{21} & 0 & a_{23} \\
			a_{31} & a_{32} & 0 	
		\end{matrix}
		\Bigg)\Bigg]_{E=E_b}
		\begin{pmatrix}
			\phi_{(1-N)} \\
			\phi_{(N+1-2N)} \\
			\phi_{(2N+1-3N)}
		\end{pmatrix}=0
	\end{gathered}
\end{align}  
In  equation (\ref{eq13}) $\phi_{(1-N)}$, $\phi_{(N+1-2N)}$, and $\phi_{(2N+1-3N)}$ are components of the algebraic wave function, projected for each partial component of two-body $t$ matrices on a momentum grid  $\mid\vec q\mid$.  It is important to note that for eigenvalue problems, the independent variables of the integral kernels $a_{ij}$ are $q^{''}$,$x^{''}$,$\phi^{''}$, and $p$, $q$, with only the last two variables being set the kinematic ones of the process and are approximated by grid momenta. That is why the solution of the system (\ref{eq13}) allows one to obtain  eigenstate  functions   projected for definite partial waves only.

The absence of a certain symmetry for the three-body wave function leads to another interesting consequence characteristic of solving systems of coupled equations. The system (\ref{eq5}) couples  different Faddeev components $\Psi_{1(23)}$, $\Psi_{2(31)}$, and $\Psi_{3(12)}$  of the total wave function between themselves. Therefore, the algebraic solutions (\ref{eq13}) must also be linearly related with each others and with  Faddeev's components, as follows from the system (\ref{eq5})
\begin{align}\label{eq14}
	\begin{gathered}
		\phi_{(1-N)}=a\big(\Psi_{2(31)}+\Psi_{3(12)}\big);\\
		\phi_{(N+1-2N)}=b\big(\Psi_{1(23)}+\Psi_{3(12)}\big);\\
		\phi_{(2N+1-3N)}=c\big(\Psi_{1(23)}+\Psi_{2(31)}\big),
	\end{gathered}
\end{align} 
where $a,b,c$ are some constants.
Since the sum of the Faddeev's components is a total wave function, which on the other hand has an algebraic approximation in the form of $\phi_{(1-3N)}$, then one leads to the relation
\begin{equation}\label{eq15}
	\phi_{(1-N)}+\phi_{(N+1-2N)}+\phi_{(2N+1-3N)}=\Psi_{1(23)}+\Psi_{2(31)}+\Psi_{3(12)},
\end{equation}
from which it automatically follows that the constants $a=b=c=1/2$. Using the relations (\ref{eq14}), it is possible to express the Faddeev's components of the total wave function of the considered system  through algebraic solutions (\ref{eq13}).
\begin{figure}[pt!]
	\begin{center}
		\resizebox{1.0\textwidth}{!}{
			\includegraphics{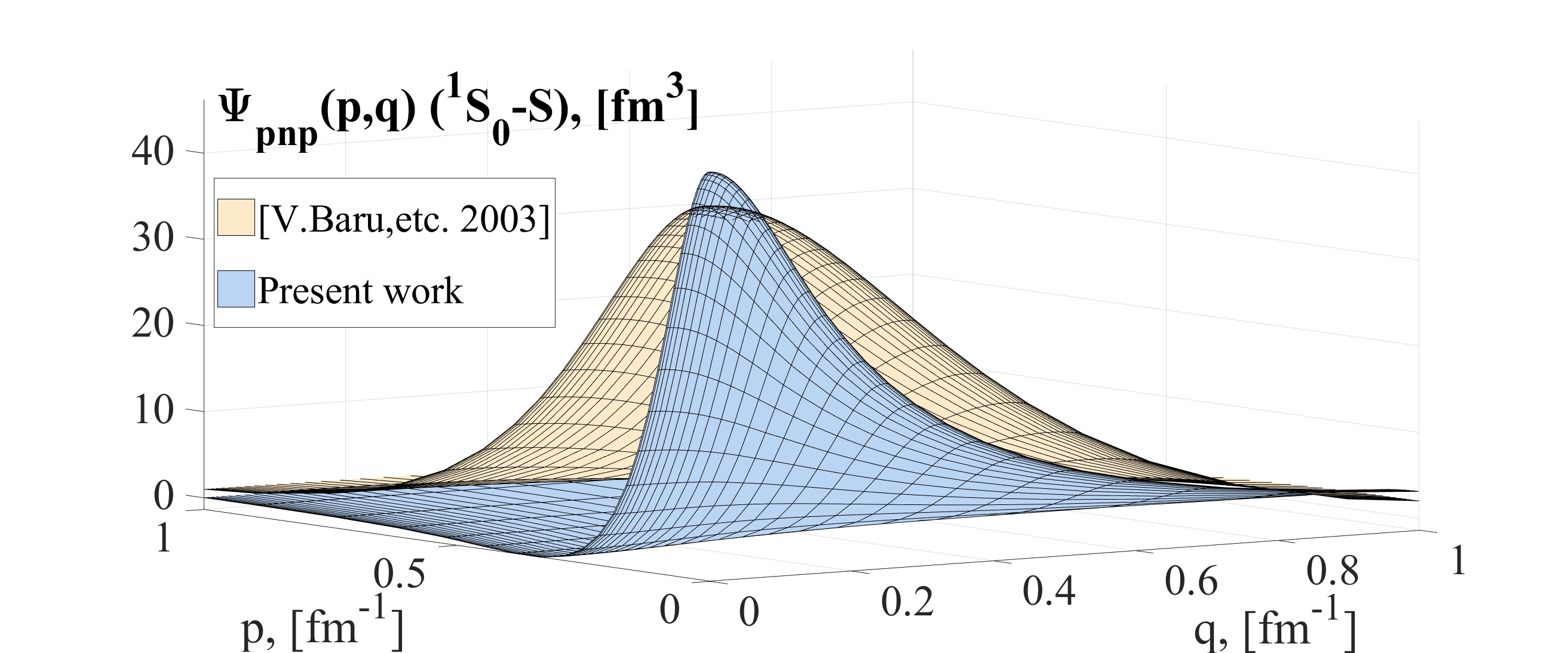}\hspace{-3cm}
			\includegraphics{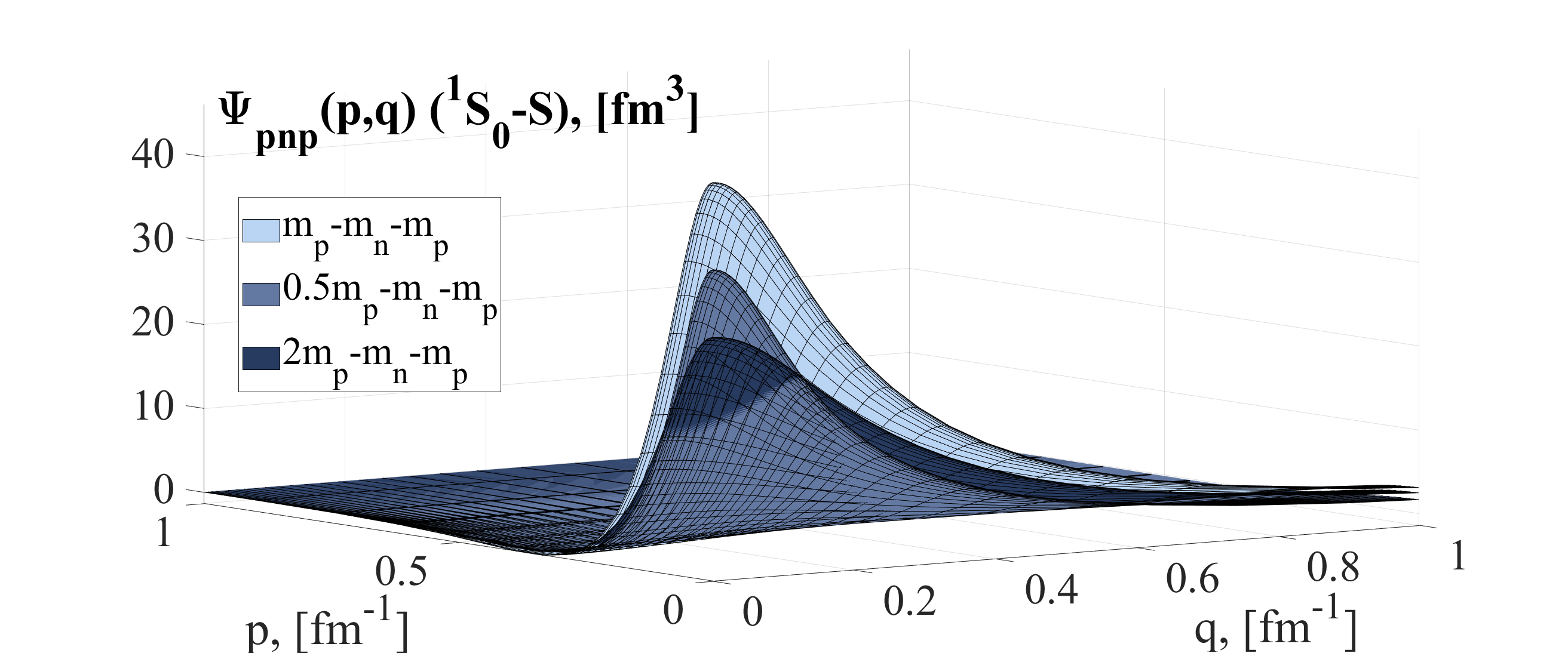}}
		\caption{Left: the partial $^1S_0-S$ component of the wave function of the three-nucleon system (p-n-p), calculated on the basis of parameterization V.Baru \cite{Baru03} and in this work using a simple Bonn separable model of NN interaction \cite{Haidenbauer86}. Right: comparison of various calculations of the Faddeev's $\Psi_{1(23)}$ component in partial $^1S_0-S$ wave  with mass variation $m_1=0.5m_p$, $m_p$, $2m_p$. }
		\label{fig1}
	\end{center}
\end{figure}		
he figure (\ref{fig1}) shows the Faddeev's components $\Psi_{1(23)}$ as an example of the wave function of the proton-neutron-proton three-nucleon system ($^3$He)  calculated at different masses $m_1=0.5m_p,m_p,2m_p$ of the spectator nucleon. The separable Bonn parameterization of the fourth rank \cite{Haidenbauer86} was used as a model of NN interaction, which provides a reliable description of the phase shifts $^1S_0$, $^3S_1-^3D_1$, as well as the binding energy of the deuteron.  For comparison, the partial $^1S_0-S$ component of the parameterization of the three-nucleon function from the work \cite{Baru03} is presented in the same figure. As one can see, despite the similar magnitude of the wave function in the region of small momenta values of the $\vec p$-interacting pair and $\vec q$-nucleon spectator, the form of attenuation of the wave function in the calculations of this work is not symmetrical and steeper than in the parameterization of \cite{Baru03}. The decrease and smearing of the wave function $\Psi_{pnp}(p,q)$ in the region of small momenta values looks unexpected with both a decrease and an increase in the mass of the $m_1$ spectator particle. 

Thus, the developed approach with direct integration of the Faddeev equations without an explicit requirement for symmetry or antisymmetry of the wave functions of interacting particles, as calculations have shown, gives not only an physically acceptable result for known three-nucleon systems with a description of their binding energies \cite{Egorov24}, but also allows one to arbitrarily change the masses of interacting particles with a corresponding change of two-body interactions. 

\section{Regions of logarithmic singularities for different particle masses}\label{sec4}

 How the regions of the logarithmic singularities of the integral kernels  $a_{ij}$   (\ref{eq5}) change depending on the masses of interacting particles?
For to answer this item, one consider  the interaction-free three-body Green function $R_0$:
\begin{align}\label{eq16}
	\begin{gathered}
		R_0=\Big(E-\frac{q^2}{2\mu_{23}}-\frac{q^{''2}}{2\mu_{13}}-
		\frac{qq^{''}y_{qq^{''}}}{m_3}\Big)^{-1};  \\
		y_0=\frac{m_3}{qq^{''}}\Big(E-\frac{q^2}{2\mu_{23}}-\frac{q^{''2}}{2\mu_{13}}\Big). 
	\end{gathered}
\end{align} 
The boundaries of the  logarithmic singularities domain are known to be determined by the equality $y_0=\pm 1$. From this simple equality, three ranges of values of the integration variable $q^{''}$ follow, outlining the area of moving logarithmic singularities as a function of the spectator momentum $\mid\vec q\mid$:
\begin{align}\label{eq17}
	\begin{gathered}
		\begin{cases}
			f_1:\,\,q^{''}=-\frac{\mu_{13}qy_0}{m_3}+\sqrt{2\mu_{13}\Big(E+q^2\Big[\tfrac{\tfrac{m_1m_2}{m_3}(y_0^2-q)-M}{2m_2(m_1+m_3)}\Big]\Big)};\\
			f_2:\,\,q^{''}=\frac{\mu_{13}qy_0}{m_3}-\sqrt{2\mu_{13}\Big(E+q^2\Big[\tfrac{\tfrac{m_1m_2}{m_3}(y_0^2-q)-M}{2m_2(m_1+m_3)}\Big]\Big)};\\
			f_3:\,\,q^{''}=\frac{\mu_{13}qy_0}{m_3}+\sqrt{2\mu_{13}\Big(E+q^2\Big[\tfrac{\tfrac{m_1m_2}{m_3}(y_0^2-q)-M}{2m_2(m_1+m_3)}\Big]\Big)}.		
		\end{cases}	
	\end{gathered}
\end{align}
Moreover, the definition area of the functions $f_1$, $f_2$, and $f_3$  for $y_0\ge 0$ are determined
\begin{align}\label{eq18}
	\begin{gathered}
		f_1\in[0,\sqrt{q_{\vee}^2}],\,f_2\in[\sqrt{q_{\vee}^2},\sqrt{q_{\wedge}^2}],\,f_3\in[0,\sqrt{q_{\wedge}^2}], 
	\end{gathered}
\end{align}
where
\begin{align}\label{eq19}
	\begin{gathered}
		q_{\wedge}^2=\frac{2m_2E(m_1+m_3)}{M-\tfrac{m_1m_2}{m_3}(y_0^2-1)};\\
		q_{\vee}^2=\frac{2m_2m_3E(m_1+m_3)}{m_1m_2+m_3M}. 	
	\end{gathered}
\end{align}
In the case when $y_0<0$, the  definition areas of the functions $f_1$ and $f_3$ (\ref{eq18}) are swapped.
\begin{figure}[pt!]
	\begin{center}
		\resizebox{1.0\textwidth}{!}{
			\includegraphics{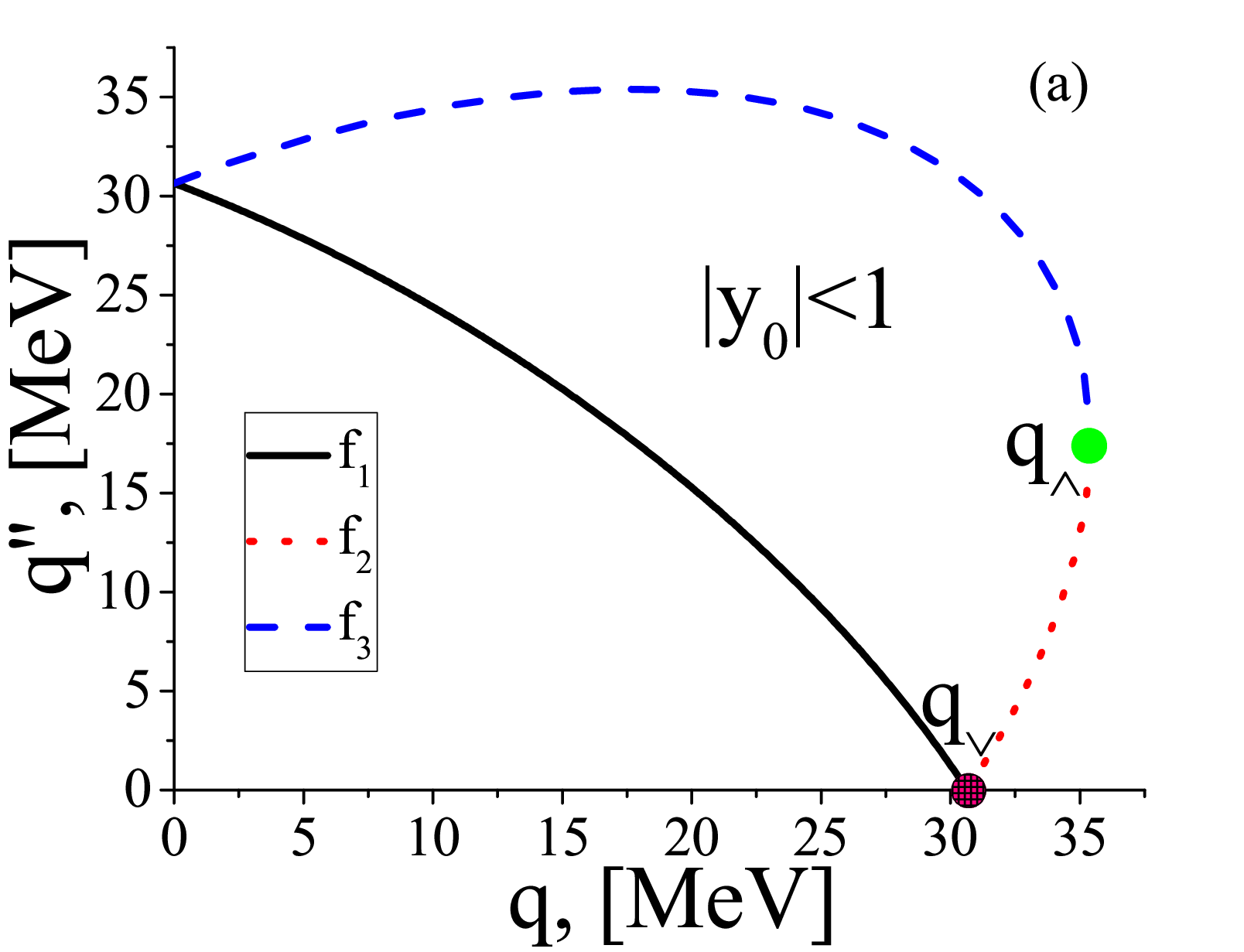}\hspace{-3cm}
			\includegraphics{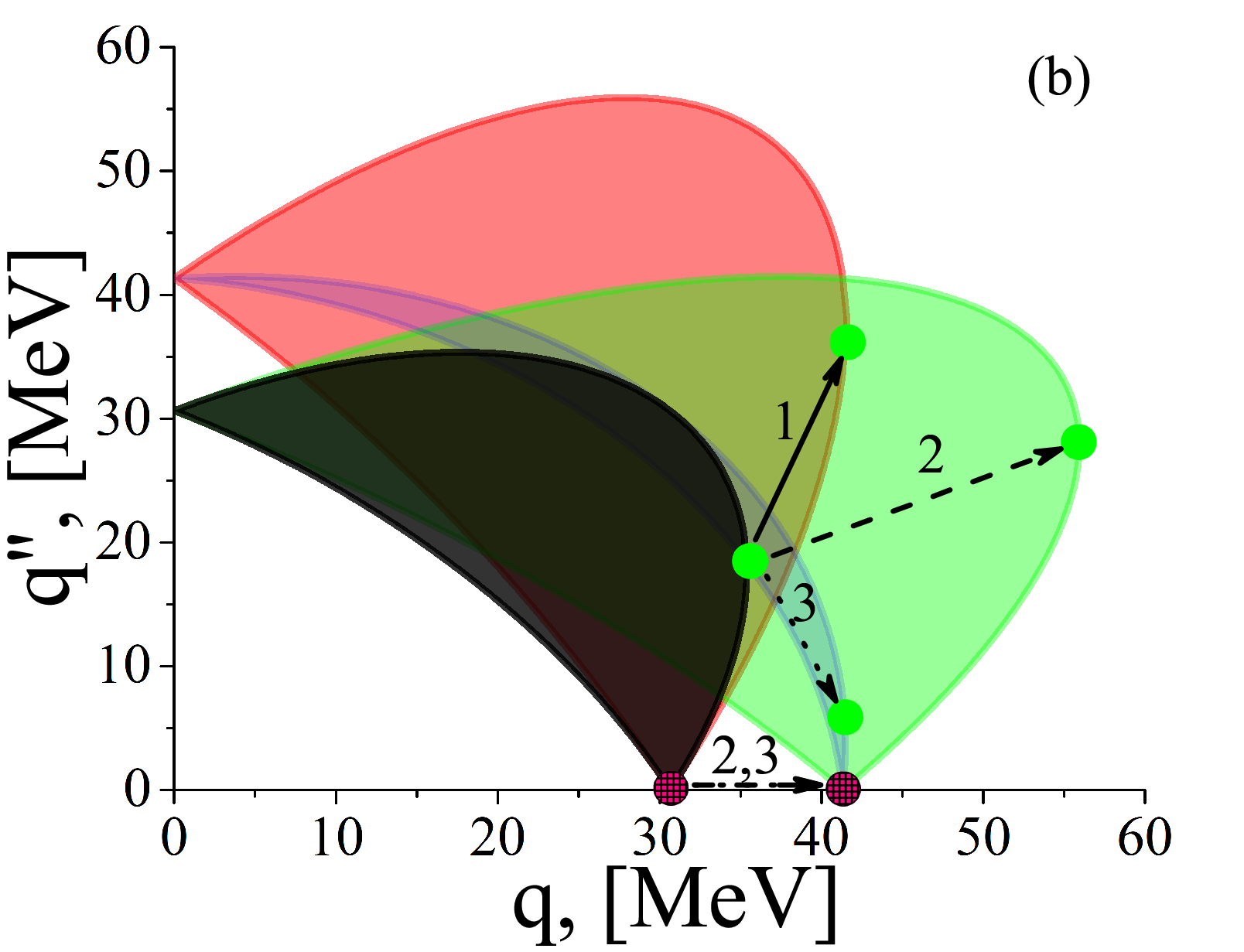}\hspace{-3cm}
			\includegraphics{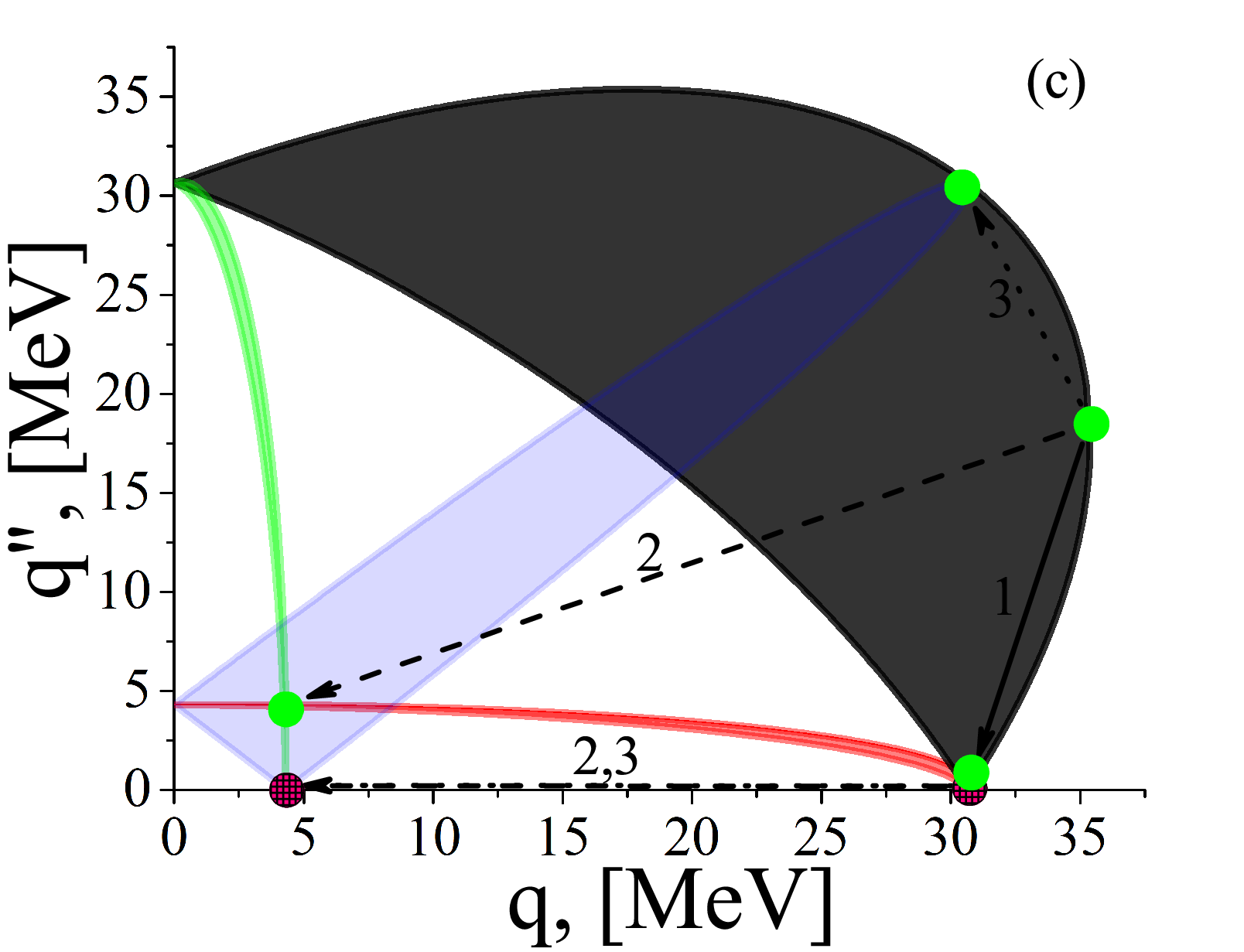}}
		\caption{Regions of logarithmic singularities at scattering energy $E=1$~MeV (a): in the system $m_p-m_n-m_p$, (b): in the systems $10m_p-m_n-m_p$, $m_p-10m_n-m_p$, $m_p-m_n-10m_p$, (c): in the systems $m_p/100-m_n-m_p$, $m_p-m_n/100-m_p$, $m_p-m_n-m_p/100$. Circles mark the positions of the boundaries of $q_{\vee}$, $q_{\wedge}$, and arrows with numbers (1,2,3 - change in mass of $m_1$, $m_2$, $m_3$, respectively) mark the movement of these boundaries from one to another system due to changes in the masses of one of the components. Additionally, the area $\mid y_0\mid<1$ is marked in Figure (a) inside the boundaries described by the functions $f_1$, $f_2$, and $f_3$ (\ref{eq17}). The colors in Figures (b) and (c) indicate the areas of logarithmic singularities: the black system is $m_p-m_n-m_p$, the red system is $10m_p-m_n-m_p$, $m_p/100-m_n-m_p$, the green system is $m_p-10m_n-m_p$, $m_p-m_n/100-m_p$ and the blue system is $m_p-m_n-10m_p$, $m_p-m_n-m_p/$100.}
		\label{fig2}
	\end{center}
\end{figure}		

The figure (\ref{fig2}) shows the regions of logarithmic singularities occurring in the proton-neutron-proton system ($m_p-m_n-m-p$ or $^3$He) at the scattering energy $E=1$~MeV. The boundaries of the integration regions (\ref{eq19}) are marked with bold dots. Changes in these boundaries due to a ten times increase and one hundred times  decrease in the mass of the $m_1$ component are marked with solid arrows with the number 1.
Similarly, boundary movements (\ref{eq19}) are denoted due to a ten times increase and one hundred times decrease in the mass of the components $m_2$ (dash-dotted arrow) and $m_3$ (dotted arrow). The zones of logarithmic singularities visible in the figure, can vary markedly depending on the change in the masses of one or another component of the system.
Increasing the energy $E$ with constant masses of the components of the system, the regions of logarithmic singularities invariably grow, while  with an increasing in the mass of one of the components, the zone can stretch along the momenta $q^{''}$ and $q$, as shown in the figure (\ref{fig2})(b), and  can also shrink into an arc as the mass of $m_3$ increases.
A similar nontrivial behavior of the  logarithmic singularities domain is seen with a sharp decrease in the mass of one of the components of the system by a factor of 100. In this case, the singularity area is transformed into an arc, and also stretches into a wedge in the case of a decrease in the mass of $m_3$ as shown in the figure (\ref{fig2})(c).
Such behavior of the  logarithmic singularities areas for systems without a certain symmetry, i.e. with a difference in particle masses, including with a strong difference of tens and hundreds of times, can play a useful role in the numerical solution of a system of inhomogeneous Faddeev equations (\ref{eq5}).
This simplification may consist in jumping over these logarithmic singularities areas in the direct numerical integration of the system (\ref{eq5}) on irregular grids of momenta $q$ and $q^{''}$ due to its smallness.

In conclusion,  one note how expressions for functions (\ref{eq17}) will change for the remaining Green's functions (\ref{eq5}). If one denote by the quantities $R_0^{[1]}$ and $R_0^{[2]}$ those Green functions that are included in the first line of the equation (\ref{eq5}), and by the quantities $R_0^{[3]}$, $R_0^{[4]}$ and $R_0^{[5]}$, $R_0^{[6]}$ those Green functions that are included in the second and third lines, respectively, then similar (\ref{eq17}) expressions can be obtained by changing the masses of particles according to the rule
\begin{align}\label{eq20}
	\begin{gathered}
		R_0\equiv R_0^{[1]}(m_1,m_2,m_3);\\
		R_0^{[1]}(m_1,m_3,m_2) = R_0^{[2]}(m_1,m_2,m_3),\,  R_0^{[1]}(m_2,m_1,m_3) = R_0^{[3]}(m_1,m_2,m_3); \\
		R_0^{[3]}(m_3,m_2,m_1) = R_0^{[4]}(m_1,m_2,m_3),\, R_0^{[2]}(m_3,m_2,m_1) =
		R_0^{[5]}(m_1,m_2,m_3); \\
		R_0^{[5]}(m_2,m_1,m_3) = R_0^{[6]}(m_1,m_2,m_3). 
	\end{gathered}
\end{align}

\section{Conclusion}\label{sec5}

In this work, inhomogeneous Faddeev integral equations for three different bodies are explicitly written out without using the traditional partial wave decomposition for the resulting breakup T-matrix. Expressions for the amplitude of elastic scattering and reaction are also written out in the context of direct integration of the obtained equations in momentum space without using partial wave decomposition.  An algebraic method for searching for eigenfunctions of stationary states of three-body systems with different masses based on homogeneous Faddeev equations is proposed and tested. The behavior of logarithmic singularities for a system of three bodies of different masses is also analyzed. It was found that at certain values of one of the masses of the three particles, these regions are compressed into an arc or elongated into a wedge-like region.

The results of this work can be directly used both for the direct solution (for example, using Pad\'e approximants) of inhomogeneous Faddeev equations for a system of three bodies of different masses to search for the scattering cross section and the reaction cross section in the cluster model of the target nucleus, and for finding various non-relativistic wave functions of three-body systems.

\backmatter
\bmhead{Acknowledgments}

Egorov Mikhail gratefully acknowledges the financial support of the 
Foundation for the
Advancement of Theoretical Physics and Mathematics “BASIS”
(project No 23-1-3-3-1).




\begin{thebibliography}{99}
\bibitem{Fadd}
L.~D.~Faddeev, Scattering theory for a three-particle system, Sov.~Phys.~JETP~{\bf 12}, 1014 (1961). 
\bibitem{FaddYakub}
O.~A.~Yakubovsky, On the integral equations in the theory of N particle scattering,  Sov.~J.~Nucl.~Phys.~{\bf 5} 937 (1967).
\bibitem{AGS67}
E.~O.~Alt, P.~Grassberger, W.~Sandhas, Reduction of the three-particle  collision problem to  multi-channel two-particle Lippmann-Schwinger equations, Nucl.~Phys.~A{\bf 2} 167 (1967).
\bibitem{AGS69}
E.~O.~Alt, P.~Grassberger, W.~Sandhas, Derivation of the DWBA in an exact three-particle theory, Nucl.~Phys.~A{\bf 139} 209 (1969).
\bibitem{Mukh12}
A.~M.~Mukhamedzhanov, V.~Eremenko, A.~I.~Sattarov, Generalized Faddeev equations in the AGS form
for deuteron stripping with explicit inclusion of target excitations and Coulomb interaction, Phys.~Rev.C{\bf 86} 034001 (2012).
\bibitem{Mukh18}
A.~M.~Mukhamedzhanov, Three-body Faddeev equations in two-particle Alt-Grassberger-Sandhas form with distorted-wave-born-approximation amplitude as effective potentials, Phys.~Rev.~C{\bf 98} 044626 (2018).
\bibitem{Fon17}
A.~C.~Fonseca, A.~Deltuva, Numerical exact ab initio four-nucleon scattering calculations: from dream to reality, Few-Body Systems {\bf 58} 46 (2017).
\bibitem{AGS70}
E.~O.~Alt, P.~Grassberger, W.~Sandhas, Treatment of the three- and four-nucleon systems
by a generalized separable-potential model, Phys.~Rev.~C{\bf 1} 85 (1970).
\bibitem{Belyaev06}
V.~B.~Belyaev, N.~V.~Shevchenko, A.~Fix, W.~Sandhas, Binding of charmonium with two- and three-body nuclei, Nucl.~Phys.~A{\bf 780} 100 (2006).
\bibitem{Shevchenko07}
N.~V.~Shevchenko, A.~Gal, J.~Mare\u{s}, Faddeev calculation of a $K^-pp$ quasibound state, Phys.~Rev.~Lett.~{\bf 98} 082301 (2007).
\bibitem{Shevchenko21}
N.~V.~Shevchenko, Four-body Faddeev-type calculation of $\bar K NNN$ system: $K^-np$ quasi-bound state, Few-Body Systems {\bf 62} 62 (2021).
\bibitem{Nogga02}
A.~Nogga, H.~Kamada, W.~Gl\"ockle, The hypernuclei $^4_{\Lambda}$H and $^4_{\Lambda}$He: challenges for modern hyperon-nucleon forces, Phys.~Rev.~Lett.{\bf 88} 172501 (2002).
\bibitem{Nogga13}
A.~Nogga, Light hypernuclei based on chiral and phenomenological interactions, Nucl.~Phys.~A{\bf 914} 140 (2013).
\bibitem{Egorov23}
M.~V.~Egorov, Searching for the bound states in the $\Xi^-nn$, $\Xi^-pn$, and $\Xi^-pp$-systems, Phys.~Atom.~Nucl.~{\bf 86} 277 (2023).
\bibitem{Mukh00}
A.~M.~Mukhamedzhanov,  E.~O.~Alt, G.~V.~Avakov, Momentum space integral equations for three charged particles: nondiagonal kernels, Phys.~Rev.~C{\bf 61} 064006 (2000).
\bibitem{Mukh01}
A.~M.~Mukhamedzhanov,  E.~O.~Alt, G.~V.~Avakov, Momentum space integral equations for three charged particles. II. Diagonal kernels, Phys.~Rev.~C{\bf 63} 044005 (2001).
\bibitem{Deltuva05a}
A.Deltuva, A.~C.~Fonseca, P.~U.~Sauer, Momentum-space treatment of Coulomb interaction in three-nucleon reactions with two protons, Phys.~Rev.~C{\bf 71} 054005 (2005).
\bibitem{Deltuva05b}
A.Deltuva, A.~C.~Fonseca, P.~U.~Sauer, Momentum-space description of three-nucleon breakup reactions including the Coulomb interaction, Phys.~Rev.~C{\bf 72} 0540044 (2005).
\bibitem{Deltuva08}
A.Deltuva, A.~C.~Fonseca, P.~U.~Sauer, Nuclear many-body scattering calculations with the Coulomb interaction, Annu.~Rev.~Nucl.~Part.~Sci.~{\bf 58} 27 (2008).
\bibitem{Deltuva19}
A.~Deltuva, A.~C.~Fonseca, P.~U.~Sauer, Coulomb force effects in few-nucleon systems, Few-Body Systems {\bf 60} 29 (2019).
\bibitem{Oryu06}
Sh.~Oryu, Two- and three-charged-particle nuclear scattering in momentum space: A two-potential theory and a boundary condition model, Phys.~Rev.~C{\bf 73} 054001 (2006).
\bibitem{Oryu06E}
Sh.~Oryu, Erratum: Two- and three-charged-particle nuclear scattering in momentum space: A two-potential theory and a boundary condition model, Phys.~Rev.~C{\bf 76} 069901(E) (2007).
\bibitem{Gloeckle96}
W.~Gl\"ockle, H.~Wita\'la, D.~H\"uber, H.~Kamada, J.~Golak, The three-nucleon continuum: achievements, challenges and applications, Phys.~Rep.~{\bf 274} 107 (1996).
\bibitem{Elster99}
Ch.~Elster, W.~Schadow, A.~Nogga, W.~Gl\"ockle, Three-body bound-state calculations without angular-momentum decomposition, Few-Body Systems {\bf 27} 83 (1999).
\bibitem{Liu05}
H.~Liu, Ch.~Elster, W.~Gl\"ockle, Three-body scattering at intermediate energies, Phys.~Rev.~C{\bf 72} 054003 (2005). 
\bibitem{Stingl}
M.~Stingl, A.~S.~Rinat (Reiner), Complete angular momentum analysis of Fadde'ev-Lovelace equations based on non-central pair forces, Nucl.~Phys.~A{\bf 154} 613 (1970).
\bibitem{Kharchenko80}
V.~F.~Kharchenko, V.~P.~Levashev, Four-nucleon problem in the integral equations approach. The binding energy of $^4$He and the n$-^3$H and n$-^3$He scattering lengths, Nucl.~Phys.~{\bf343} 249 (1980). 
\bibitem{Kilic04}
S.~Kilic, J.~Ph.~Karr, L.~Hilico, Coulombic and radiative decay rates of the resonances of the exotic molecular ions $pp\mu$, $pp\pi$, $dd\mu$, $dd\pi$, and $dt\mu$, Phys.~Rev.~A{\bf 70} 042506 (2004). 
\bibitem{Kolganova11}
E.~Kolganova, A.~K.~Motovilov, W.~Sandhas, The $^4$He trimer as an Efimov system, Few-Body Systems {\bf 51} 249 (2011).
\bibitem{Frolov19}
A.~M.~Frolov, On the weakly-bound (1,1)-states in the three-body $dd\mu$ and $dt\mu$ muonic ions, Phys.~Rev.~A{\bf 383} 1288 (2019).
\bibitem{Kolganova19}
E.~A.~Kolganova, V.~Roudnev, Weakly bound LiHe$_2$ molecules in the framework of three-dimensional Faddeev equations, Few-Body Systems {\bf 60} 32 (2019). 
\bibitem{Orlov81}
Yu.~V.~Orlov, On analytic continuation of integral equations of the scattering theory to a nonphysical sheet of energy, Pisma~Zh.~Eksp.~Teor.~Fiz.~{\bf 33} 380 (1981).  
\bibitem{Orlov84}
Yu.~V.~Orlov, Investigation of few-baryon systems with the use of  integral equations [in russion],  Doctors Thesis, Nucl.~Phys.~Inst.~of~Moscow~State~Univ., 1984. 
\bibitem{Baru03}
V.~Baru, J.~Haidenbauer, C.~Hanhart, J.~A.~Niskanen, New parameterization of the trinucleon wave function and its application to the $\pi^3$He scattering length, Eur.~Phys.~J.~A{\bf 16} 437 (2003). 
\bibitem{Haidenbauer86}
J.~Haidenbauer, Y.~Koike, W.~Plessas, Separable representation of the Bonn nucleon-nucleon  potential, Phys.~Rev.~C{\bf 33} 439 (1986). 
\bibitem{Egorov24}
M.~V.~Egorov, Binding energies of $^3$H, $^3$He nuclei in three-body Faddeev equations with direct integration, Phys.~of~Atom.~Nucl.~{\bf 87} 682 (2024). 
\end{thebibliography}

\end{document}